\documentclass[sigconf]{acmart}

\usepackage{xcolor}
\definecolor{DarkGreen}{HTML}{000000}
\definecolor{orange}{HTML}{000000}

\newcommand\rr[1]{\textcolor{DarkGreen}{#1}}

\AtBeginDocument{%
  }

\copyrightyear{2025}
\acmYear{2025}
\setcopyright{cc}
\setcctype{by}
\acmConference[CHI '25]{CHI Conference on Human Factors in Computing Systems}{April 26-May 1, 2025}{Yokohama, Japan}
\acmBooktitle{CHI Conference on Human Factors in Computing Systems (CHI '25), April 26-May 1, 2025, Yokohama, Japan}\acmDOI{10.1145/3706598.3713262}
\acmISBN{979-8-4007-1394-1/25/04}

\begin{document}

\title{Understanding Children's Avatar Making in Social Online Games}


\author{Yue Fu}
\email{chrisfu@uw.edu}
\orcid{0000-0001-5828-5932}
\affiliation{%
  \department{Information School}
  \institution{University of Washington}
  \city{Seattle}
  \state{Washington}
  \country{USA}
}

\author{Samuel Schwamm} 
\email{Samuel.Schwamm@childrens.harvard.edu}
\orcid{0009-0007-1956-6820}
\affiliation{%
  \department{Digital Wellness Lab}
  \institution{Boston Children's Hospital}
  \city{Boston}
  \state{Massachusetts}
  \country{USA}
}

\author{Amanda Baughan}
\email{baughan@cs.washington.edu}
\orcid{0000-0003-2217-6974}
\affiliation{%
  \department{Allen School of Computer Science}
  \institution{University of Washington}
  \city{Seattle}
  \state{Washington}
  \country{USA}
}

\author{Nicole Powell} 
\email{Nicole.Powell@childrens.harvard.edu}
\orcid{0009-0002-7175-1162}
\affiliation{%
  \department{Digital Wellness Lab}
  \institution{Boston Children's Hospital}
  \city{Boston}
  \state{Massachusetts}
  \country{USA}
}


\author{Zoe Kronberg} 
\email{zoe.kronberg@childrens.harvard.edu}
\orcid{0009-0004-2029-5931}
\affiliation{%
  \department{Digital Wellness Lab}
  \institution{Boston Children's Hospital}
  \city{Boston}
  \state{Massachusetts}
  \country{USA}
}


\author{Alicia Owens} 
\email{alicia.owens@childrens.harvard.edu}
\orcid{0009-0004-7112-1472}
\affiliation{%
  \department{Digital Wellness Lab}
  \institution{Boston Children's Hospital}
  \city{Boston}
  \state{Massachusetts}
  \country{USA}
}

\author{Emily Izenman}
\email{Emily.Izenman@childrens.harvard.edu}
\orcid{0000-0002-4687-4024}
\affiliation{%
  \department{Digital Wellness Lab}
  \institution{Boston Children's Hospital}
  \city{Boston}
  \state{Massachusetts}
  \country{USA}
}

\author{Dania Alsabeh} 
\email{alsabehd@umich.edu}
\orcid{0009-0006-5120-8321}
\affiliation{%
  \department{Department of Pediatrics}
  \institution{University of Michigan}
  \city{Ann Arbor}
  \state{Michigan}
  \country{USA}
}

\author{Elizabeth Hunt} 
\email{ehunt@edc.org}
\orcid{0000-0002-2714-1124}
\affiliation{%
  \department{Digital Wellness Lab}
  \institution{Boston Children's Hospital}
  \city{Boston}
  \state{Massachusetts}
  \country{USA}
}

\author{Michael Rich} 
\email{Michael.Rich@childrens.harvard.edu}
\orcid{0000-0001-8721-3516}
\affiliation{%
  \department{Digital Wellness Lab}
  \institution{Boston Children's Hospital}
  \city{Boston}
  \state{Massachusetts}
  \country{USA}
}

\author{David Bickham} 
\email{David.Bickham@childrens.harvard.edu}
\orcid{0000-0002-2139-6804}
\affiliation{%
  \department{Digital Wellness Lab}
  \institution{Boston Children's Hospital}
  \city{Boston}
  \state{Massachusetts}
  \country{USA}
}

\author{Jenny Radesky} 
\email{jradesky@umich.edu}
\orcid{0000-0002-7721-7350}
\affiliation{%
  \department{Department of Pediatrics}
  \institution{University of Michigan}
  \city{Ann Arbor}
  \state{Michigan}
  \country{USA}
}

\author{Alexis Hiniker}
\email{alexisr@uw.edu}
\orcid{0000-0003-1607-0778}
\affiliation{%
  \department{Information School}
  \institution{University of Washington}
  \city{Seattle}
  \state{Washington}
  \country{USA}
}

\renewcommand{\shortauthors}{Fu et al.}

\begin{abstract}
Social online games like Minecraft and Roblox have become increasingly integral to children's daily lives. Our study explores how children aged 8 to 13 create and customize avatars in these virtual environments. Through semi-structured interviews and gameplay observations with 48 participants, we investigate the motivations behind children's avatar-making. Our findings show that children's avatar creation is motivated by self-representation, experimenting with alter ego identities, fulfilling social needs, and improving in-game performance. In addition, designed monetization strategies play a role in shaping children's avatars. We identify the ``wardrobe effect,'' where children create multiple avatars but typically use only one favorite consistently. We discuss the impact of cultural consumerism and how social games can support children's identity exploration while balancing self-expression and social conformity. This work contributes to understanding how avatar shapes children's identity growth in social online games.

\end{abstract}




\maketitle

\section{Introduction}
In 2023, over 3.38 billion people worldwide play video games \cite{Sarah_2023}. A report shows more than 90\% of children older than 2 years play video games in the USA, and three-quarters of American households own a gaming console \cite{alanko2023health}. For Generation Alpha—those born after 2010—82\% of boys and 76\% of girls play video games for at least one hour per week \cite{essential}, and 43\% of children aged 8-12 report playing video games every day \cite{rideout2021common}. \rr{Social online games} like Minecraft, Fortnite, and Roblox, which offer multiplayer experiences and encourage social interactions, have become some of the top downloaded and played games in history. By June 2024, Minecraft alone has amassed 166 million active players worldwide, with 80\% of children playing alongside friends, family, and other online users \cite{minecraftStats}. In these virtual worlds, children collaborate on creative projects, explore vast landscapes, defend bases, and tackle challenges together, making these games a central part of their social and recreational lives.

One of the key features of these social games is the ability for players to create and customize avatars. Avatars act as the digital representation of the player, serving as the primary means for interaction, communication, and engagement in games, offering a medium for self-expression and identity experimentation. Understanding how and why children customize their avatars is crucial, as avatar can affect player's presence and arousal \cite{bailey2009avatar, chung2008avatar}, engagement \cite{castronova2003theory}, cognitive and emotional processing \cite{bailey2009avatar}, identity formation \cite{li2013player, villani2016exploration}, motivation for learning \cite{mazlan2012students}, sense of control \cite{foster2008games}, and overall game enjoyment \cite{turkay2010enjoyment, chung2008avatar, bailey2009avatar, Birk2016, trepte2010avatar}.  For children, the customization of an avatar provides a unique opportunity to explore different facets of their identity and preferences in a virtual environment. Research indicates that children's avatars both reflect and shape their developing sense of self, encompassing factors such as age, gender, race, and socio-economic status \cite{villani2016exploration}. However, this exploration is often limited by social norms and game design, which can reinforce stereotypes, such as the sexualization of female avatars \cite{crowe2014click}, and lack of availability of racial minority characters \cite{fields2012navigating}. Additionally, children's avatar customization choices are frequently influenced by in-game and real-world currency and purchase design \cite{hota2016real}, which are often designed to encourage spending.

The influence of avatar creation and customization on self-identity and presentation is likely to be particularly significant among younger users, who are in the critical stages of identity development and are more socially self-conscious than adults. Despite this, little research has been conducted on what motivates children to create and customize their avatars. Our study examines children's motivations for avatar creation in social games like Minecraft and Roblox. We aim to answer the following questions:

\begin{itemize}
    \item\textbf{RQ1:} What motivates children to create and customize their avatars in social games?
    \item\textbf{RQ2:} How, if at all, does the design of \rr{monetization mechanisms in game} shape this process?
\end{itemize}

To explore these themes, we conducted a study with 48 participants between the ages of 8 and 13 who regularly play Roblox, Minecraft, Fortnite or other social online games. We conducted semi-structured interviews about participants' experiences playing social online games while they played a game of their choice. Our investigation focuses on how children create and customize their avatars and how game design and monetization mechanisms impact children’s avatar creation and customization. We analyzed both the visual characteristics of the children's avatars and their self-reported motivations for avatar creation by reviewing video recordings of gameplay and their responses during the interviews.

Our findings show that children are motivated to create and customize avatars in social games for various reasons: to accurately represent themselves, to experiment with alter ego identities, to connect socially with friends and family, and to enhance in-game performance. Additionally, monetization mechanisms encourage children to invest in paid customization options, leading to what we identify as the "wardrobe effect"---accumulating multiple avatars but consistently using only a favorite one.

This study contributes to a understanding of how avatar creation impacts children's identity development in virtual environments. By highlighting the interplay between children's motivations, game design, and monetization mechanisms, we provide insights for designing social games that support identity exploration while balancing self-expression with social conformity, and the influence of culture consumerism. \rr{In addition, we contribute an anonymized dataset of 80 participants' avatar visuals (see Fig. \ref{fig:avatar sample} in Appendix for a sample)}. Our work offers implications for game designers to foster healthy digital experiences for children.

\section{Related Work}
\subsection{Children's Social Online Game}
Social games typically provide a gaming experience that involves more than one player, emphasizing interactions between players and spectators. Social game interactions can include voice or text-based chat, non-verbal communication like pointing and touching, message boards, and features such as combat and trading \cite{DeKort,Emmerich,GONCALVES2023107851}. Although there are various definitions of social games, one review paper identified five common ways that research characterizes social gaming: 1) non-solitary play, (2) design intent, (3) the interactions it promotes, (4) the resulting social outcomes, and (5) the inherent social nature of gaming. Social games often challenge players in ways that promote competition and collaboration, requiring active communication both within and outside the game environment to achieve success \cite{Depping2018,depping2017cooperation,Depping2016,nardi2006strangers}.

\rr{Social online games are multiplayer games where players interact with each other through computer mediated communication and collaborative or competitive gameplay \cite{domahidi2014dwell, domahidi2018longitudinal}.} These games are immensely popular among children and adolescents, who make up the majority of users. For example, Minecraft has become the best-selling game of all time, with over 180 million monthly active players \cite{minecraftStates}. Similarly, Roblox, another prominent social and Metaverse gaming platform, has experienced dramatic growth, reaching over 70 million daily active users worldwide \cite{robloxStats}. In 2024, Roblox reported over 32 million daily active users under the age of 13, up from 28.2 million in the previous year for the same age group \cite{robloxStats-children}. These platforms have become central to the Metaverse experience for Gen Z (born from 1995 to 2009) and Gen Alpha (born from 2010 to 2024) children.

Playing digital social games is crucial for children's social well-being and the development of social-emotional skills \cite{parkash2022utilizing,Granic2014-aj,przybylski2010motivational}. Young people often play with peers, discuss games, and use gaming as a way to expand their social interactions and relationships, thereby increasing social closeness \cite{lenhart2008teens,Orleans2000, olson2010children, depping2017cooperation}. Research highlights the cognitive, motivational, emotional, educational, health, and social benefits of gameplay \cite{Granic2014-aj}. For instance, for children on the autism spectrum, games can serve as powerful tools for social and meta-cognitive learning \cite{Bartoli2013, Hiniker2013}, with Minecraft, in particular, empowering autistic youth to modify the game to better support their emotional and social needs \cite{Ringland2016}. Co-playing with parents and other children can also serve as a bonding opportunity, with studies finding that increased gaming time with siblings leads to greater affection \cite{coyne2016super,COYNE2011160}.

Despite the recognized benefits of social games, parental opinions are mixed. Concerns include children spending excessive time on video games \cite{mcclain2022parents}, games interfering with family time or sleep \cite{Freed2020}, and exposure to violent or sexual content \cite{kutner2008parents}. From the perspective of young players, while they acknowledge the risks associated with social gaming—such as playing with a mean-spirited stranger\cite{Carter2020}—many view these social features as integral to their enjoyment of the game \cite{livingstone2021playful}. However, the value of sibling play in social gaming remains understudied \cite{kaitlin2024}.

In our study, we allowed parents and children to determine their interpretation of social gaming and interviewed children who play Minecraft, Roblox, Fortnite, Genshin Impact, and Rocket League. Our focus was on exploring children’s perspectives on avatar creation and the motivations and meanings behind their choices in these social games.

\subsection{Avatars in \rr{Social Online Game}}
Avatars aren't merely digital puppets; they become sites of 'subjectification,' where individuals are both shaped by and actively negotiate systems of governance. Avatar customization choices are influenced by platform affordances, social norms, and community expectations, but users also creatively express themselves through these constraints, sometimes even subverting them \cite{bardzell2014lonely}. In social online games, players can create and customize avatars to serve both as their visual representation and the primary means of interacting with the virtual world and other players \cite{szolin2022gaming, castronova2003theory, mazlan2012students}. An avatar can be any element that represents a user, from human-like characters to non-human representations such as cars, animals, or even icons and text \cite{boberg2008designing}, and most avatars combine elements of both reality and imagination \cite{mazlan2012students}. Social online games often offer extensive customization options, allowing users to modify various aspects of their avatars, such as hairstyles, skins (the textures that are placed on avatars, e.g. in Fortnite), outfits, accessories, and non-human features. When creating avatars, players often balance their self-image with how they want to be perceived by others \cite{szolin2022gaming}. Avatars are central to social online games, enabling players to express their identities, preferences, and personalities in a virtual environment. The ability to customize avatars has been shown to influence players’ sense of presence \cite{bailey2009avatar}, playtime \cite{castronova2003theory}, cognitive and emotional processing \cite{bailey2009avatar}, emotional arousal \cite{chung2008avatar},  and overall enjoyment \cite{turkay2010enjoyment, bailey2009avatar, Birk2016, trepte2010avatar}. 

Factors such as the capacity to identify with a person or a character motivates players to engage in avatar creation \cite{foster2008games}. The identification process allows players to experiment with different aspects of their personalities and strategically select traits they wish to project through their avatars \cite{waggoner2007passage, triberti2017changing, wood2020me}. \rr{For example, research shows gifted adolescents may identify as ``gamers,'' and features for customizing avatars offer them opportunities for exploring multiple identities, experimenting with different roles, and navigating social interactions in a low-risk environment \cite{wood2020me}. Studies about avatars in Virtual Reality (VR) show that while some users experiment with different identities through their avatars, many prioritize presenting a consistent self that resembles their physical self. Aesthetics play a significant role in initial social interactions, with visually appealing avatars more likely to draw positive attention \cite{freeman2021body}.} In addition, identifying with avatar can even influence players' behavior outside the game \cite{yee2007proteus}, making them more susceptible to persuasive messages \cite{moyer2011identification}.

Prior research Research shows people’s avatars in online games both reflect and contribute to aspects of their identity formation, including aspects of age, gender, race/ethnicity, and socio-economic status \cite{triberti2017changing}. For example, some researches find young people use avatars to explore gender by simply presenting as a different gender or developing gender-diverse identities in preparation for coming out as transgender or non-binary in real life \cite{crowe2014click, morgan2020role, hussain2008gender}. In addition, another research notes as adolescents get older, they create more detailed representations of themselves, including features associated with changes during adolescence and puberty, showing avatars are reflections of players real-world change. \rr{The avatar-user relationship is not monolithic but complex and dynamic, varying across different virtual worlds and contexts.  The relationship between a user and their avatar(s) evolves over time, reflecting changing uses, purposes, and expectations \cite{de2012my}.} 

\rr{Compared to adults, children’s avatar creation is often more focused on playfulness and creativity. Children are likely to choose avatars based on fun and imaginative aspects, often opting for characters that are colorful, whimsical, or fantastical \cite{kriglstein2013study}. This difference in approach can be attributed to cognitive development stages; children are still forming their self-concept and may not yet engage in the same level of self-reflection as adults \cite{principe2013children}. In addition, children do not exhibit the same level of concern for gender representation in avatars as adults. Their choices are often influenced by the characters they admire from media or their peers, leading to a more fluid approach to gender in avatar selection \cite{kriglstein2013study}. }

Game design can affect how players create avatars. For instance, Crowe \& Watts \cite{crowe2014click} found that avatar customization options for female characters in a game were often stereotypical and highly sexualized, featuring exaggerated body shapes and revealing clothing. The option limitation influences players variably, while some teenage girls expressed frustration with this limited representation, others appreciated the ability to make their characters appear conventionally attractive. Similarly, another study shows there are limited customization options for avatars that do not align with the default racial characteristics often associated with White players \cite{fields2012navigating}. 

\rr{Monetization strategies in social online games also play a crucial role in shaping avatar customization behaviors. Micro-transactions—small in-game purchases made with real-world money or virtual currency—enable players to acquire digital goods, upgrades, or content that enhance their avatars \cite{gibson2022relationship}. These transactions range from purely cosmetic items, such as skins that alter the appearance of characters or items, to more functional enhancements that can impact gameplay. Research has demonstrated that monetization mechanisms exploit various cognitive biases to drive repeat purchasing. For example, loot boxes, pay-to-win models, and limited-time offers leverage the sunk-cost fallacy and scarcity bias to encourage players to spend more frequently \cite{james2022between, petrovskaya2021predatory, gonzalez2024mediating}. The frequency and nature of these micro-transactions have been identified as key indicators of problematic gaming and gambling behaviors \cite{gibson2022relationship}.}

Specifically relating to avatar customization, micro-transactions foster the development of virtual shopping behaviors among children aged 8-12. Hota and Derbaix \cite{hota2016real} observed that children engage in purchasing accessories to personalize their avatars, with motivations varying by gender. Boys are often driven by the desire for functional advantages related to power and progress within the game, while girls are more motivated by the need to enhance their social status and aesthetic appeal. \rr{These digital items are inherently alluring and socially significant, compelling children to invest in paid options to improve their social standing and identity representation in online spaces \cite{james2022between}. Moreover, predatory monetization strategies exacerbate these behaviors by obscuring real-world costs and creating barriers to spending transparency. In-game virtual currencies often make it difficult for players, especially children, to understand the true monetary value of their purchases \cite{james2022between, petrovskaya2021predatory}. These strategies frequently result in bundled sales, where players accumulate surplus virtual currency that cannot be easily spent, thereby encouraging further purchases to utilize these leftover funds \cite{report}.}

Our study builds upon this literature by focusing on children's perceptions and motivations in avatar creation and how game design and monetization influences children's choices. While previous research has explored the role of avatars in games mostly in the view of adults, we contribute to this field by examining how children perceive, create, and customize avatars, and how game design and monetization strategies shape these processes. By interviewing children, we provide insights into their needs and motivations in avatar-making within digital social games.

\section{Method}

\rr{This study was a multi-institutional effort involving researchers from three U.S. sites: the Adolescent Medicine division of a pediatric hospital (Site A) and two major research universities (Sites B and C). Site A served as the lead institution and obtained Institutional Review Board (IRB) approval for the study. The Principal Investigators (PIs) at Sites B and C and their respective IRBs signed reliance agreements through the SmartIRB platform. The PIs included two pediatricians with extensive clinical research experience and two university faculty members with doctoral degrees in [details anonymized].}

\rr{Research assistants (RAs) at all three sites collaborated on recruitment, data collection, and analysis. To maintain consistency, RAs underwent comprehensive training sessions covering study procedures, consent/assent protocols, and the semi-structured interview guide. Each RA completed a mandatory practice session with a volunteer participant before conducting live sessions. Additionally, a detailed project manual was provided to all RAs, outlining every step of the study process—from recruitment through data collection and analysis—to ensure consistent implementation across the three institutions.}

\subsection{Participants}

To understand how children create avatars in social games and how the gaming environment facilitates and influences these creations, we conducted an observational and semi-structured interview study with 51 children in which child participants played a social online game with a research assistant observed and asked questions. We recruited participants from three U.S. institutions using a combination of flyers, email lists, and parent research pools. \rr{We considered a ``social online game'' as any online multiplayer video game that enables player-to-player interaction beyond solitary play, including communication and collaborative or competitive gameplay. The study was advertised as a video game study to understand children's gameplay, experiences, and perceptions. We provided examples of social online games such as Minecraft, Roblox, and Fortnite and mentioned that other social online games are welcome.  We listed these three titles since Roblox, Minecraft, and Fortnite are particularly popular in this age group and well-known for their social features. Children who play other social online games are likely to play one of these three. In addition, advertising these titles also facilitated the research team's preparedness, as RAs are required to be familiar with the game terminology and basic gameplay mechanics. In practice, most participants signed up to play these three titles, with two other participants playing Genshin Impact and Rocket League, which were also included in the dataset.} 

Parents filled out an eligibility form, which asked for the child's age, gender, games they play, frequency of play, parent contact information, and zip code. To be eligible, children needed to be aged 8-14, fluent in English, and play online multiplayer video games at least once a week. \rr{This age range was chosen to capture a broad spectrum of young users. Participants at this age are generally capable of articulating their gameplay experiences and preferences, with older participants (11-14 years old, including those approaching early middle school age) more likely to provide valuable insights into how peer dynamics shape social interactions in gaming environments.} Interviews and gameplay sessions were conducted via Zoom, requiring participants to have compatible devices where they could share their gameplay during the interview. 

We received in total 286 eligibility surveys entries (some of which were duplicated or incomplete). We screened children for eligibility, prioritizing recruiting a diverse sample, including girls and participants of color, who are historically underrepresented in U.S. studies of children’s technology use. 

We ended up recruiting and interviewing 51 participants (21 from site A, 15 from site B, and 15 from site C). Three interviews were excluded from analysis due to poor audio quality or technical problems during the interview. In total, we analyzed 48 interview recording sessions. Our sample had a median age of 10 years old consisting of 28 boys, 18 girls, and 2 transgender/non-binary individuals. The racial/ethnic distribution was as follows: 23 White, 12 multi-racial/ethnic (i.e. selected more than one category), 7 Asian, 4 Black/African American, 1 Hispanic/Latino, and 1 “other”.

Participants played a variety of social games of their choice during the interview, including Roblox (29), Minecraft (13), Fortnite (4), Genshin Impact (1), and Rocket League (1). In our initial eligibility survey, parents reported that participants play with friends/peers (N=37), by themselves (N=35), with siblings (N=20), and/or with parents/caregivers (N=7) . The duration of their experience with the game of choice varied: 8 participants had been playing for five or more years, 13 for 3-4 years, 20 for 1-2 years, and 7 for less than a year.

\subsection{Procedures}

After screening for eligibility, researchers from the three institutions organized a 20-30 minute phone call with the participant and their parent to introduce the study, discuss risks and benefits, and review the consent form. At end of the call, the research assistant (RA) scheduled a 90-minute gameplay Zoom interview session with the family \rr{for a later date (typically within one week)}. During the scheduled interview, the RA started by reviewing the consent form and study details again with the family, answering questions, and obtaining verbal assent from the child and electronic consent from the parent via REDCap, a data management system for research \cite{patridge2018research}.

The interview was comprised of two parts. In the first part, the RA asked the child to tour the game according to a semi-structured protocol, asking the child to visit the game's home page, describe their avatar and username, review their friends list, and visit the game’s marketplace. In the second part of the interview, the child engaged in free play, during which the RA asked questions related to the ongoing game play activities on the screen. Children were instructed to play as they usually would but to avoid interacting with other players to protect their privacy. The interviews were recorded using Zoom, capturing audio of the child’s responses and video of the child’s shared screen showing the game. Only the RA's face and the gameplay screen were recorded; the child's username and face were not captured.

The data analyzed here are part of a larger investigation into children’s social gaming. In the current study, we focused only on the responses to the questions in our protocol that related to avatars. We asked questions such as: 

\begin{itemize}
    \item Why did you choose that [avatar] person/hair/outfit/weapon/face etc.? 
    \item How do you feel about the options that are available?
    \item What do you like about this outfit? Would you wear it in real life? 
    \item How often do you change your avatar/outfit? 
    \item What do you think the way your avatar looks says about you? 
    \item How do you set yourself apart from others? Or do you like to go unnoticed? 
    \item Can you show me the stuff you’ve bought? Why did you buy that?
    \item Do you usually act like yourself when you’re playing, or do you take on a different personality?
    \item Do you change how you play the game when you change your skin/avatar?
\end{itemize}
After the interview, we thanked the child and parent and compensated them with a \$25 gift card. The interviews were transcribed using Zoom's native transcription feature, with additional accuracy checks via a Python script and manual editing. The study was approved by the IRBs of the collaborating institutions.

\subsection{Analysis}

We conducted a thematic analysis \cite{braun2012thematic} of participants' responses to the avatar-related interview questions and a visual analysis of their avatars. Six research assistants extracted relevant segments from the transcripts and watched the screen share videos at timepoints where the participants discussed their avatar. \rr{Each RA 1) identified segments of transcripts related to avatar characteristics, customization processes, and rationales; 2) captured representative screenshots of avatars and customization interfaces; and 3) took initial notes of their observations and interpretations. The RAs then organized these extracted segments, images, and notes on a shared Miro board \cite{miroboard}, grouping data by participant ID. This visual and textual arrangement allowed the team to see each child’s avatar(s) and game contexts alongside their verbalizations (see Fig \ref{fig:miro coding}  in Appendix)}. 

\rr{The team met bi-weekly (and occasionally weekly) online. During these meetings, RAs shared their observations and notes, commented on emerging themes, and discussed how the data related to the research questions. Our approach to theme development was primarily inductive, allowing patterns to emerge organically from the data. Nonetheless, we remained aware of existing literature on children’s development, game design, and social interactions in digital contexts, which sensitized us to certain concepts. PIs from both pediatric and children's technology backgrounds participated in the meetings and shared their expertise and understanding.
As we advanced, initial codes capturing children’s motivations (e.g., desire for self-representation or experimentation) and contextual factors (e.g., game design, monetization strategies) were iteratively grouped into higher-level categories. We then revisited and refined the initial themes and resolved disagreements through group discussion. One RA reviewed all transcripts, applying the themes to quotes in a shared group document. In addition, the RA reorganized the document based on the consensus after each meeting.}

\rr{In parallel, two RAs organized all relevant avatar screenshots into a visual image board for closer examination (see Fig. \ref{fig:coding image board} in Appendix). Each team member reviewed the visuals by adding the notes next to specific visuals. This enabled us to compare visual patterns (e.g., style, accessories, complexity) against children’s stated preferences and their avatar presentations. After several rounds of coding and refinement, one RA synthesized the themes and supporting quotes into a document, which was used to write the Results section below.}

\section{Results}
\begin{figure}[h] 
    \centering
    \includegraphics[width=1\linewidth]{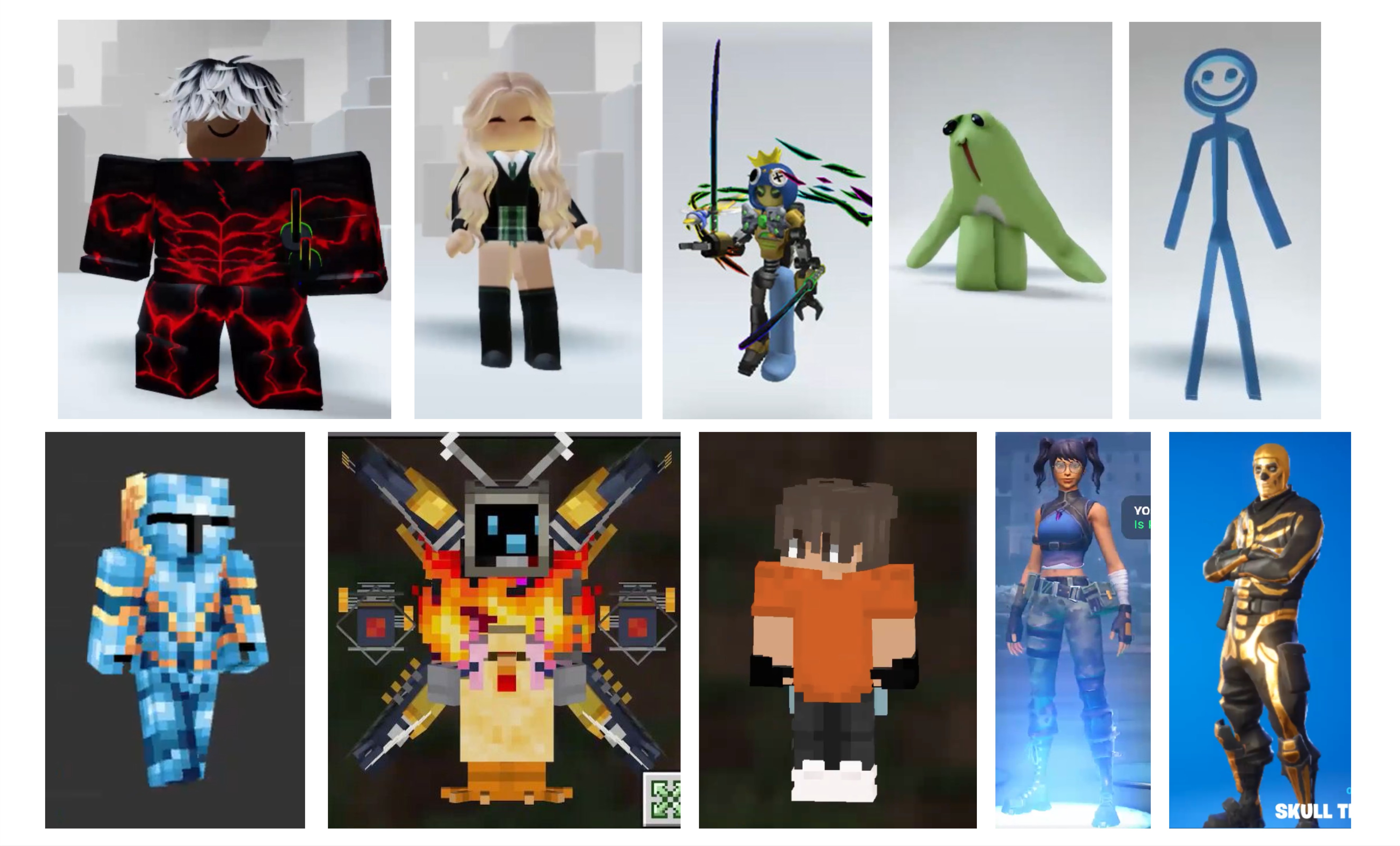} 
    \caption{An image board displaying diverse avatars created by children: the top row features Roblox avatars, while the first three on the bottom left are from Minecraft, and the rest are from Roblox.}
    \label{fig:avatar board}
\end{figure}

The ability to customize avatars in these games allows children to experiment with ideas and representations that are often wild and imaginative. The flexibility offered by social games encourages them to explore a wide range of designs, from minimalist to extravagant or even bizarre. This customization process reflects the broad spectrum of creativity and diverse preferences among children (see Fig\ref{fig:avatar board}). 

Children have the ability to customize their avatars in various online games, each offering distinct mechanisms and features for this purpose. Most social online games provide a centralized hub where players can create, purchase, and visualize their avatars, a process frequently discussed by children during our interviews. To help readers better understand these terms and their context, we present a brief introduction to key avatar-related terms and their meanings in three of the most popular social games: Roblox, Minecraft, and Fortnite.

\begin{figure}[h] 
    \centering
    \includegraphics[width=1\linewidth]{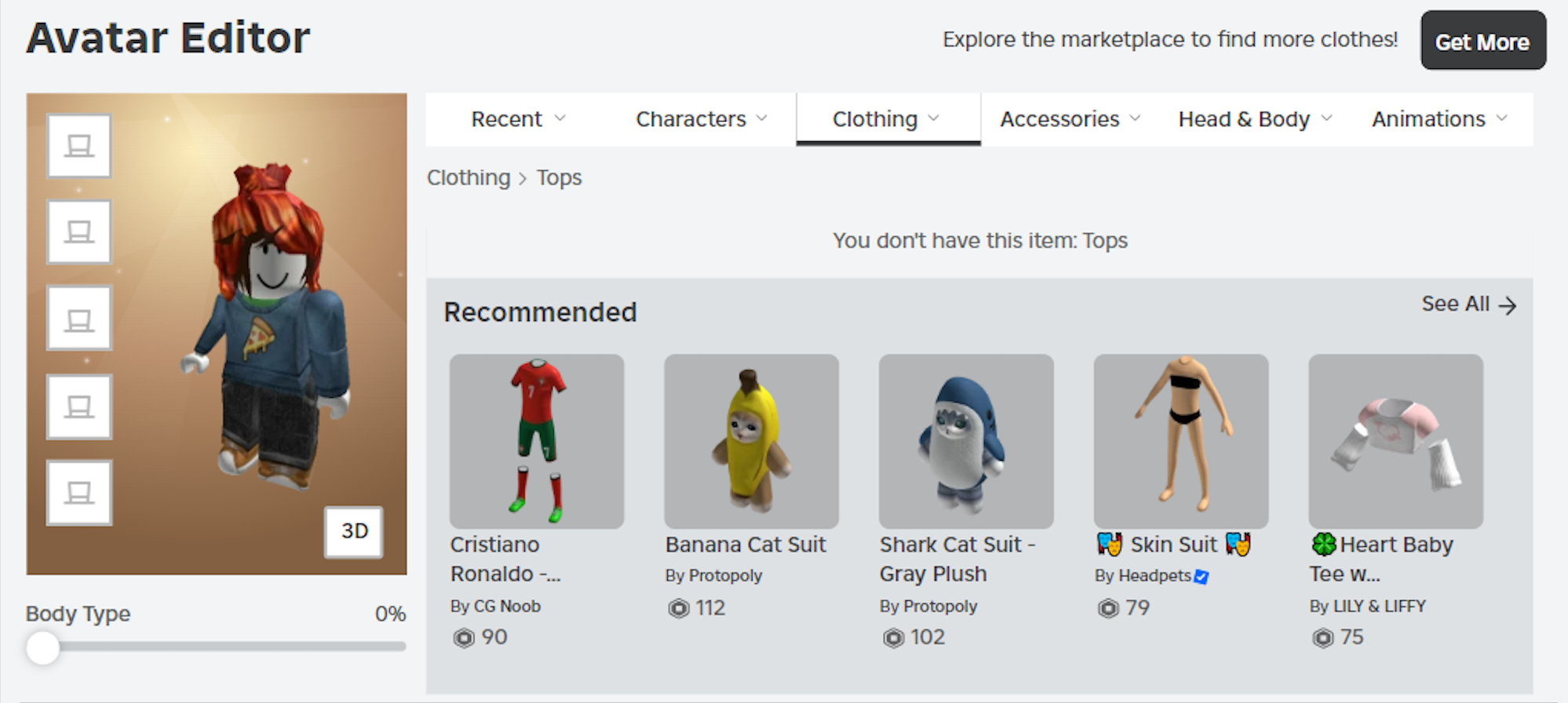} 
    \caption{Avatar customization in Roblox's Avatar Editor. Children can customize their characters, clothing, accessories, head \& body, and animation. Recommended items usually cost Robux (shown as a small coin icon bottom of items), the in-game currency.}
    \label{fig:roblox_avatar_editor}
\end{figure}

In \textit{Roblox}, players use the \textit{Avatar Editor} (see Fig. \ref{fig:roblox_avatar_editor}) to construct and modify their avatars by selecting from a diverse array of body parts, clothing, accessories, and animations. Additional items can be acquired through the \textit{Avatar Shop} using the in-game currency, \textit{Robux}. \textit{Minecraft} offers avatar customization differently across its editions: in the Java Edition (tailored for PC users who seek extensive modding capabilities and customization), players upload custom skin files via the \textit{Minecraft Launcher}, allowing for personalized designs, while the Bedrock Edition (for non-PC platforms) provides an in-game \textit{Character Creator} for modifying body features and attire, with options available for purchase using \textit{Minecoins}. \textit{Fortnite} enables avatar personalization through the \textit{Locker}, where players can equip owned skins and cosmetic items like back blings and harvesting tools. The currency in Fornite is called \textit{V-Bucks}.

\subsection{The Motivations Behind Children's Avatar Customization}
We found that the visual appearance of children's avatars varied widely, reflecting a diverse mix of facial expressions, hair styles, outfits, and accessories (such as headphones, hats, swords, and more). At times, children described simple visual appeal as the main driver behind their choices. They said things like, ``\textit{I realized that these shoes and pants looked really cool, so I chose that},'' (P31) and ``\textit{`I just thought they were pretty, so I decided to get them}'' (P33). In some instances, children described searching having a particular style holistically, explaining that ``\textit{I really like spooky stuff}'' (P30), ``\textit{I usually want stuff that would look sick, like superhero stuff}'' (P71), or describing a preference for ``\textit{medieval stuff}'' (P155) or ``\textit{odd skins that don't really resemble a human}'' (P135).

However, children's descriptions of the motivations behind their avatar-creation decisions also went deeper than simple visual preference. Specifically, children described using avatar-construction as a way to: 1) accurately portray their physical self, 2) experiment with an alter ego, and 3) connect socially with others. We describe each of three types of work below.

\subsubsection{An Opportunity to Accurately Portray Themselves}
\begin{figure}[h] 
    \centering
    \includegraphics[width=1\linewidth]{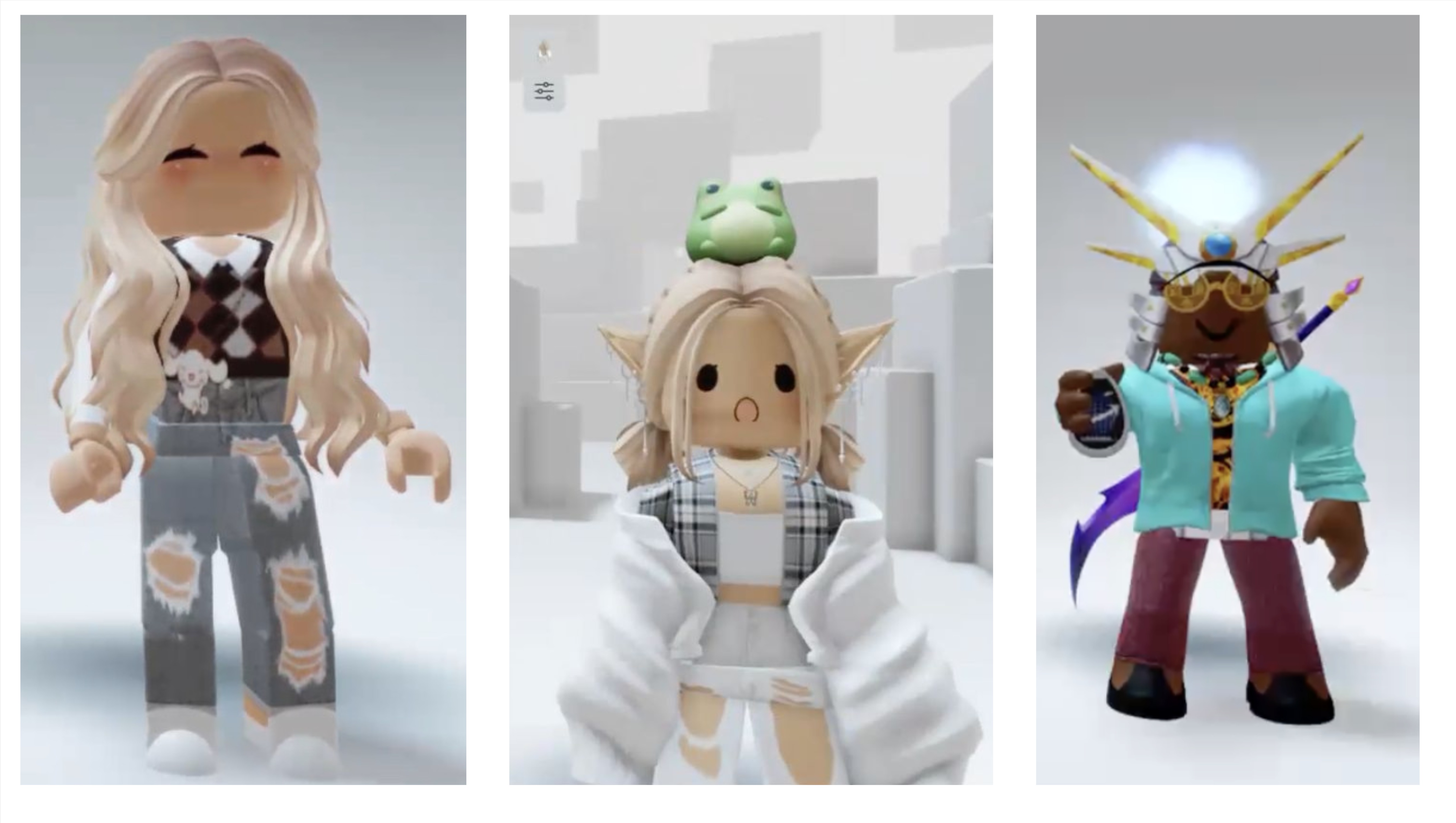} 
    \caption{Avatars created by three child participants to experiment with alter ego. From left to right: Avatars by P111, P140, and P12.}
    \label{fig:portraying self}
\end{figure}

Many children chose to create avatars that closely resemble their real-life appearance (see Fig. \ref{fig:portraying self}). These participants expressed a desire to match their avatars' features with their own, including eye color, hair color, clothing choices, race and ethnicity, thereby creating a digital representation that mirrors their physical identity. For example, one participant mentioned, ``\textit{It has black hair like me. I mean, I liked it. It was the first thing right here}'' (P59). Similarly, another child explained, ``\textit{You can see I chose a hair that's closer to my own}'' (P33), showing a preference for avatars that can represent them accurately. One participant detailed their efforts to replicate their appearance, saying, ``\textit{I tried to make it as close to my own one as I can remember, so I made it the same hat}'' (P124), underscoring the importance of having access to avatar designs that reflect physical characteristics children identify with. These examples and others suggest that at least some children value the ability to create avatars that are consistent with their appearance in the offline world.

We also found that children used their avatars to project aspects of their personalities. When asked what their avatar's outfit said about them, one child replied, ``\textit{That I'm kind of like a happy person. And that I like flowers}'' (P62). Another child mentioned that their avatar's appearance shows ``\textit{that I'm sweet. 'Cause it's supposed to have a face that smiles}'' (P108). These quotes illustrate ways in which children use avatars to project internal aspects of themselves in addition to external, physical attributes.
\begin{figure}[h] 
    \centering
    \includegraphics[width=1\linewidth]{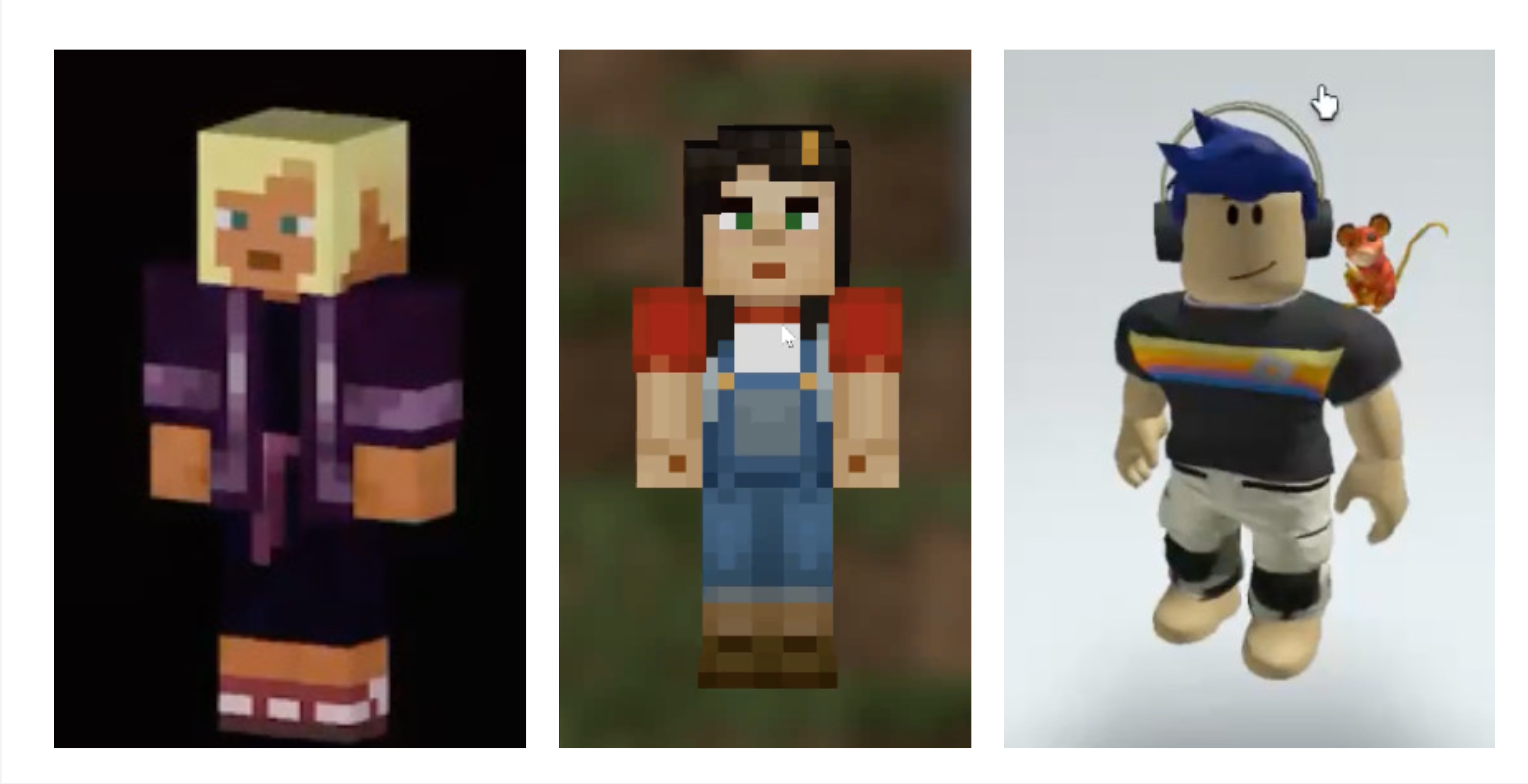} 
    \caption{Avatars created by three child participants to represent themselves. From left to right: Avatars by P250, P59, and P222.}
    \label{fig:alter_Ego}
\end{figure}

\subsubsection{A Chance to Experiment with an Alter Ego}
However, we also consistently saw that children adjusted the details of their avatars in ways that differed from or went beyond the child's offline self (see Fig\ref{fig:alter_Ego}). Many children made adjustments to their avatar as a way to experiment with styles they would never try out in the physical world. This included adopting characteristics they do not currently possess but aspire to have, or adding accessories that would violate social norms or other constraints in the physical world. In these cases, children's avatars mostly resemble the child’s physical self, but with enhancements that reflect their fantasies and aspirations.

For example, one participant described creating an avatar that matched their daily appearance but added elements they would not normally wear: ``\textit{I don't really wear ripped jeans, but it just looks good on the avatar}'' (P111). Another child selected a suit and tie for their avatar, explaining that they would only be able to wear such an outfit themselves in specific contexts, like ``\textit{when I go to church, or have a special occasion}'' (P135). When asked if they would wear their avatar’s outfit in real life, one child responded, ``\textit{Probably not! The glasses and the sword would be okay. I am not sure about the helmet}'' (P12, the avatar had a unique and large helmet). Another child explained that their avatar's hairstyle ``\textit{doesn't really match my hair\ldots the avatar’s [hairstyle] looks like a grown-up, because \ldots I would like be more independent}'' (P140). In these and many other instances, children began by creating an avatar that looks like them, but then used the freedom of the gaming environment to experiment with more imaginative, unconventional, or aspirational looks that they could not wear in the real world.

\subsubsection{A Chance to Connect Socially with Others}
Children also used their avatar choices as a way to connect socially with others, both in real life and within virtual gaming environments. For example, many children create avatars that reflect the preferences and styles of their family members. One participant shared, ``\textit{I really like this one because it kind of looks like my dad when I'm playing normal Minecraft}'' (P43). Another described choosing a blue avatar because ``\textit{my dad's favorite color is blue}'' (P187). Children also choose avatars to match those of their friends as a way to foster community and shared identity. For instance, one child mentioned, ``\textit{my friend also has this Slytherin outfit, the Harry Potter one. So we matched on that one}'' (P111). Others expressed similar sentiments, saying things like, ``\textit{one of the people I know from summer camp used this skin}'' (P9), and ``\textit{I just based it off my friend; she used to use this face}'' (P81). These quotes highlight how children customize their avatars as a way of reinforcing social bonds.

In some cases, children even alter their avatars based on friends' requests or to provoke reactions from friends. One participant explained:

\begin{table}[H]
\begin{tabular}{r p{.72\columnwidth}}
\textbf{Interviewer}:& ``\textit{Yeah. What would make you wanna change it}?''
\\\textbf{P81}:& ``\textit{If I was getting bored of it or if my friends told me to dress up for a game or something}.''
\\\textbf{Interviewer}:& ``\textit{Do you and your friends dress up together}?''
\\\textbf{P81}:& ``\textit{Sometimes in games}.''
\\\textbf{Interviewer}:& ``\textit{So what kind of outfits do you guys dress up in when you guys are dressing up together}?''
\\\textbf{P81}:& ``\textit{I just like cheap shirts so we can find it in the marketplace}.''
\end{tabular}
\end{table}
\noindent Another child explained, ``\textit{It just kinda looks silly and funny. And I wanted to see my friend's reaction\ldots me and my friend were doing a cat-themed world in Minecraft, and I just kinda picked this out}'' (P36).

Separately, children explained that customizing their avatar also gives them a way to attract social attention and stand out in the gaming community. One participant described their avatar helping them achieve social recognition, saying, ``\textit{He [a player in game] came back because I'm cool. And he wants to be with cool people\ldots Some people [come back] if you have a cool avatar.}'' (P111). Children also mentioned that they seek feedback and appreciate compliments on their avatars from other players in game. As one child put it, ``\textit{Yeah, I'm like, `Rate the drip [avatar],' then they say `10 out of 10.' People do like it. Even my friends at camp say I have a cool One Piece [a popular manga series] skin}'' (P68). Participants described their desire to attract attention via their avatars by saying things like, ``\textit{I think I'd like my skin to stand out more than camouflage}'' (P155). And others described explained that in gaming culture, unique or expensive avatars function as status symbols. As one child observed, ``\textit{If something's really expensive on Roblox, of course they're gonna show\ldots I think they flex 'cause maybe they're just happy that they really got this cool item}'' (P68). Another child echoed, ``\textit{they just want to be different}'' (P184), explaining that avatars can be used to attract attention and recognition within the gaming community.

We observed that children who primarily play alone tended to place less emphasis on customizing and decorating their avatars compared to those who are more socially engaged in multiplayer settings. For instance, one child (P3) who played a game centered around a single individual jumping off cliffs and winning points by breaking bones showed little interest in customizing their avatar. In contrast, another child (P50) who played a game where they took on the role of a baby being cared for by others demonstrated more interest in avatar customization, as the game involved more social interaction, even if only with in-game characters.

\subsubsection{A Promise of Improved In-Game Performance}
\begin{figure}[h] 
    \centering
    \includegraphics[width=0.3\textwidth]{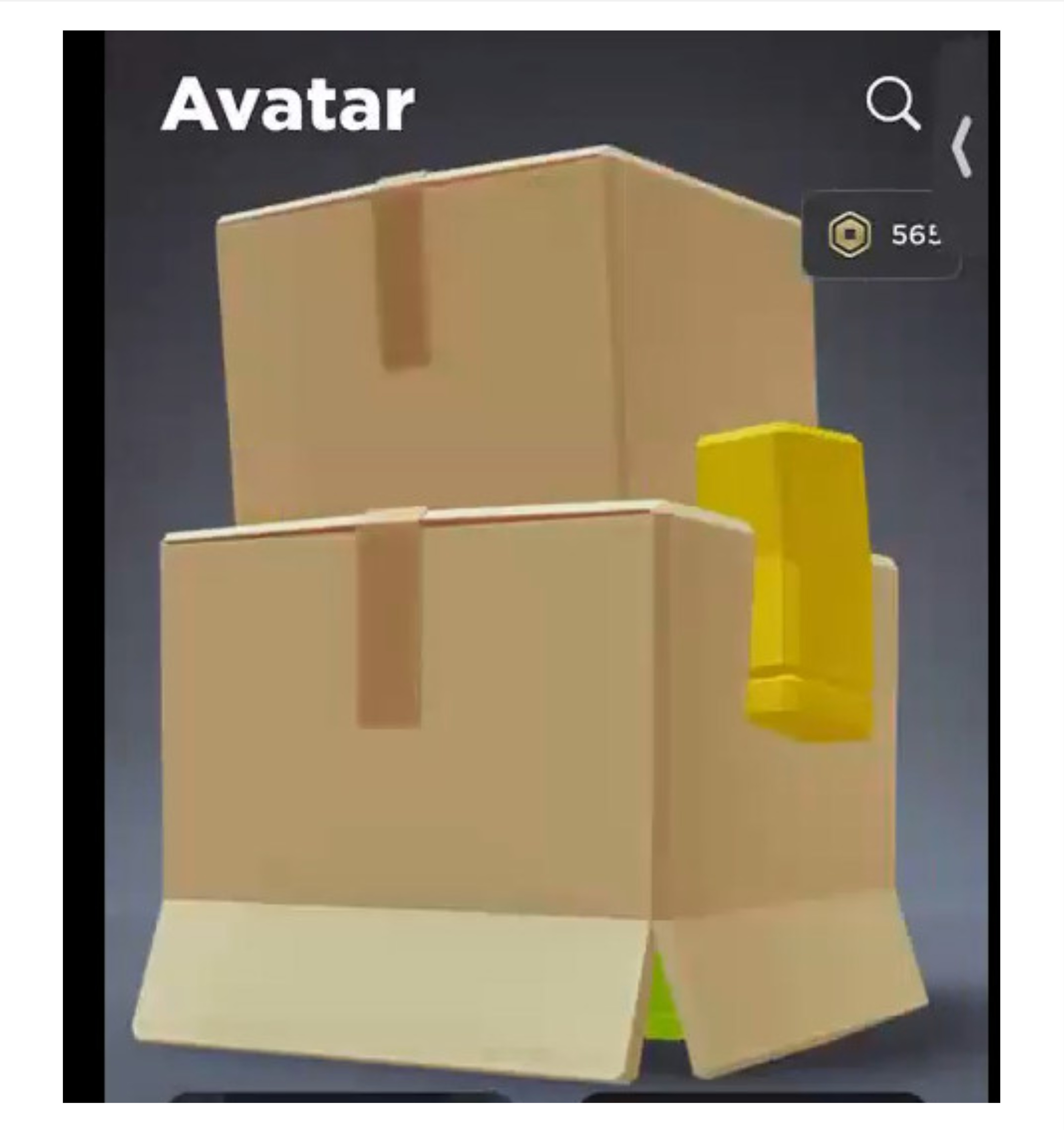} 
    \caption{Participant 81 felt the avatar could improve their camouflage and strength.}
    \label{fig:promised function}
\end{figure}

Finally, children described customizing their avatars and adding accessories in hopes of improving their performance in the game. In some instances, these hopes were realistic, and children found that they could adjust their avatars to improve their \textit{camouflage} and \textit{strength}. For example, one participant explained that their previous avatar had been a stacked box, because ``\textit{I thought it would be like a cool disguise in some games, like a few years ago, but I don't play them anymore}'' (P81). Another added that they choose customizations ``\textit{that look really cool, they blend in with the surroundings, like camouflage}'' (P155). The design of environments and how avatars interact with them likely reinforces these beliefs. See Fig \ref{fig:promised function}  for an example.

In other instances, children described customizing their avatars to make them stronger. They explained these mechanics, saying things like, ``\textit{they're kits that you can choose and they help you in the game and you can equip them; some, you buy with Robux, some are free for a week}'' (P79). Another child, discussing the Roblox game ``Murder Mystery,'' noted, ``\textit{the big[ger] your character is, the bigger the hitbox is for the murderer and for the arm sheriff\ldots The bigger it is, the harder it gets for them to kill you}'' (P68).

However, in many instances these enhancements consisted more of style than substance, and it was often unclear if they delivered the performance improvements children sought. We saw that in-game marketplaces routinely display and advertise avatar accessories in ways that suggest these customizations might enhance gameplay, when in actuality, their impact is minimal or nonexistent.
In response, children often select items that have the appearance of improving performance or offering strategic benefits. For instance, one child described their avatar’s elaborate customization, saying:
\begin{quote}
    ``\textit{I'm thinking of him as like this guy that has TNT protection, and he's on fire or got magic or something. His stuff is kinda like an air mask. His top is a lava shirt, so that it'll make that kind of fiery, sparky effect. He's got lava pants because that also makes even more fiery, sparky stuff. His outerwear is an explosion-proof set.” }However, when asked if these customizations had a real impact, the child admitted\textit{, “Unfortunately, it doesn't change how he's affected in the game}'' (P68).
\end{quote} This was consistent with the responses of other participants, who also expressed an interest in cosmetic enhancements with no tangible effect.

\subsection{How Designed Monetization Mechanisms Influence Avatar Making}
\begin{figure}[h] 
    \centering
    \includegraphics[width=1\linewidth]{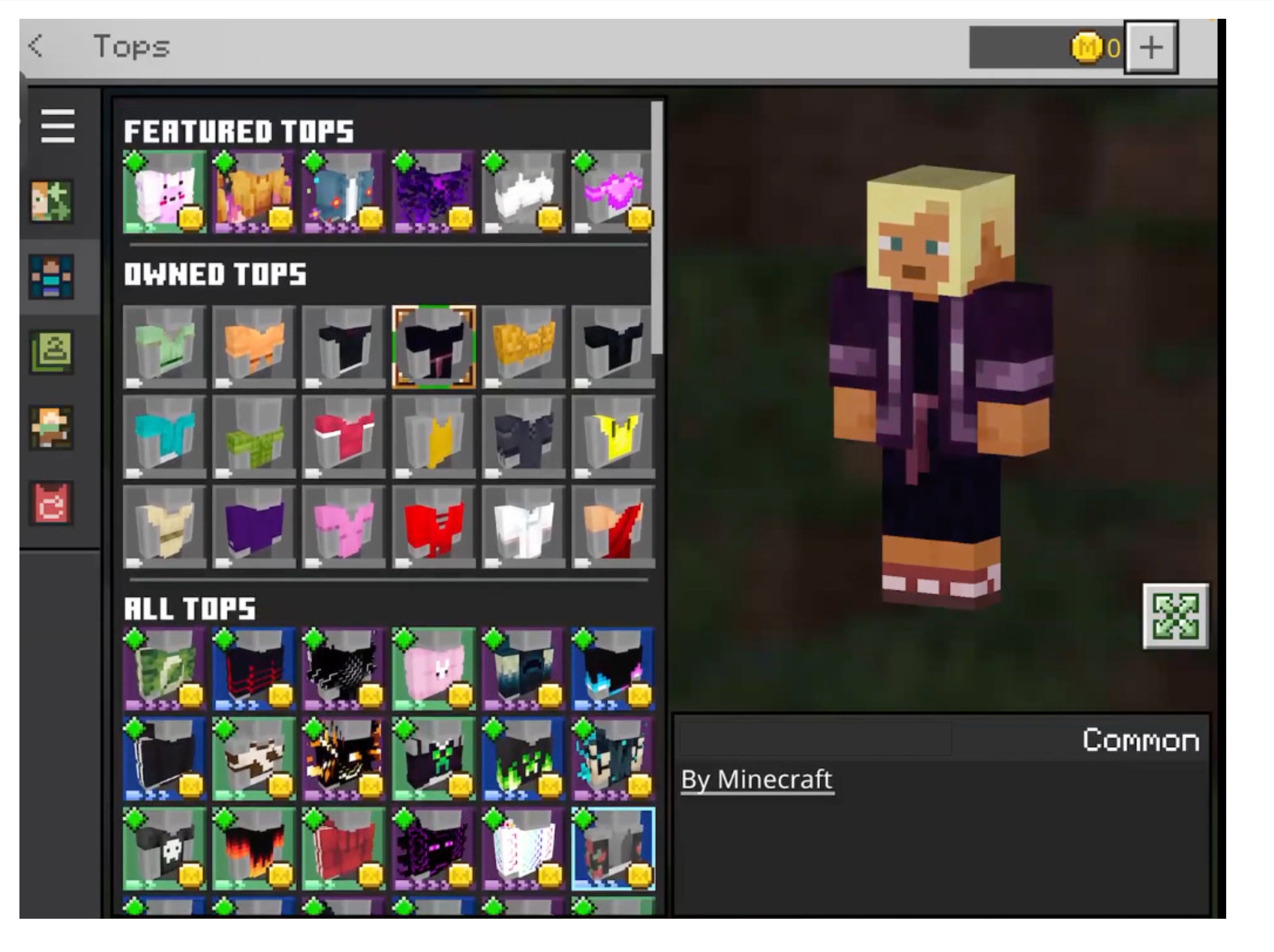} 

    \caption{Minecraft avatar customization interface, displaying various options. The top-left section shows "Featured Tops," and the lower rows display "All Tops," both of which include paid items that are more visually appealing and highly decorated. In contrast, the middle section labeled "Owned Tops" consists of free items that are simpler and lack decorative elements.}
    \label{fig:paid or free}
\end{figure}

Children's behaviors and responses revealed a great deal about how avatar customizations are monetized to consumers. Children explained that they acquire and customize avatars through three main methods. First, they described having access to free options, which generally serve as a starting point for customization. Second, they described purchasing avatars and customizations using in-game currency like ``V-Bucks'' or ``Robux.''
Third, they described purchasing game passes or bundles which provide many options (e.g., ``\textit{The Battle Pass has a lot of things, and it costs 900 V-Bucks. These singular skins cost way more, and the Battle Pass also gives you more V-Bucks than it costs}'' (P155)). Here, we describe children's perspectives on these free and paid options and the surrounding design mechanisms that encourage purchases (see Fig. \ref{fig:paid or free}).

\subsubsection{Free Avatars: Unappealing Options that Signal Lack of Experience}
free avatars are frequently viewed as simple and ``\textit{boring}'' (P84). Children often associate default skins with ``\textit{newbies},'' (P40) which can carry a negative social status within the game. One participant expressed their reluctance to use default skins, saying, ``\textit{I didn't really want to use any of the [default] skins\ldots like these skins at the bottom, because when I first got the game, they were just too simple}'' (P126). Another participant elaborated on the social implications, explaining, ``\textit{people make fun of them, not just because of the bacon hair, but because it probably means they're newer}'' (P108). Similarly, another child compared a cute, paid hairstyle with the basic hair, noting, ``\textit{[the paid ones] are really cute fun like a little pigtail hair. And on the less expensive side, there's like what they call bacon hair}'' (P133). This association between default skins and novice status reinforces a desire to purchase more distinctive and visually appealing avatars.

\subsubsection{Paid Avatars: The Appeal of Customization and Status.} Children are drawn to paid avatars and decorations because these items often offer more elaborate and visually appealing options. As one participant noted, ``\textit{I think it's fun to customize your avatar and be like, `Oh, my God, this is so pretty. I want to buy it.' So in games, most of the time, the people who spend a lot of money are the ones who spend it on clothing. That's like the most common thing you'll spend it on}'' (P33). Another participant mentioned, ``\textit{If it looks really nice and good, then it probably cost Robux}'' (P12). And a third described splurging on a fancy outfit, saying, ``\textit{The helmet was free, but the clothes were not. I bought those with the 5,000 Robux I had}'' (P71). This willingness to spend large amounts of in-game currency highlights the value children place on uniqueness and customization, which the games actively promote by offering aesthetically pleasing and desirable options.

\begin{figure}[h] 
    \centering
    \includegraphics[width=1\linewidth]{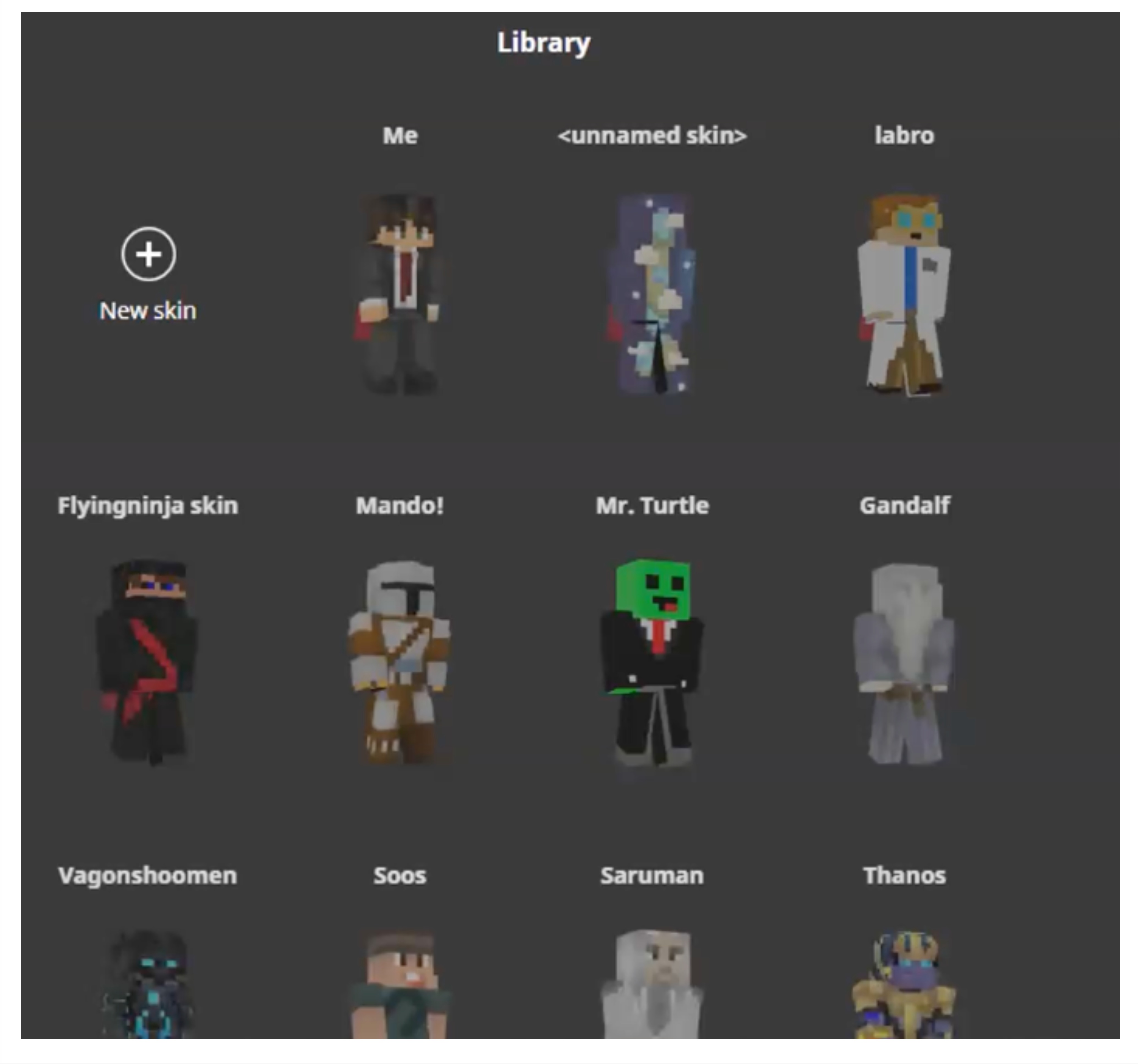} 
    \caption{A child's avatar library featuring a variety of branded characters, such as Gandalf, Saruman, and Thanos, reflecting the influence of popular media and entertainment franchises on children's avatar choices.}
    \label{fig:branded}
\end{figure}

\subsubsection{The Allure of Branded and Promoted Content}
Children described being drawn to culturally relevant content. They explained that they customize their avatars to reflect ideas and content from both entertainment media and social media. In games that allow players to purchase branded characters to use as avatars, children said they often choose avatars based on what they find appealing or what is trending at the moment (see Fig. \ref{fig:branded}. As one participant explained, ``\textit{I made my avatar based on something I watch or because it's from something that's popular right now}'' (P81).

Many children reported that they select avatars that reflect their favorite characters from books, TV shows, and movies. For instance, one child said, ``\textit{I like [the avatar]. Last season of Fortnite, they did a Peter Griffin update}'' (P250), showing how avatar characters featuring cultural icons can influence choices. Another participant expressed their love for various franchises: ``\textit{Gandalf, I love Lord of the Rings, and the Mandalorian. This is a character from Gravity Falls, which is a TV show I really like. Saruman, who's again from Lord of the Rings. And then Thanos, who's from the Avengers. I just kinda thought it was cool}'' (P135). Other children mentioned a variety of other popcultural characters, such as those from Jurassic World (P30), Harry Potter (P111), One Piece (P68), and Among Us (P30). Some explained that they construct these avatars themselves, saying, for example, ``\textit{I have an avatar of [popular TV-show character] Walter White. I think I got all of it [the outfit] for free except for the face, the Walter White face}'' (P81). These choices reflect the deep connection children have with the media they consume and how it translates into their virtual identities.

Social media platforms, particularly YouTube and TikTok, also play a significant role in influencing children's avatar choices. Many children mentioned that they draw inspiration from watching YouTubers or other players' gameplay videos. For example, one child explained, ``\textit{Mainly if I see this YouTube video, if it's one of my favorite characters, I'll try to make them in Roblox}'' (P68). Another participant shared, ``\textit{I was watching YouTubers just use it, so I just got it}'' (P81). Others described drawing inspiration from TikTok, saying, ``\textit{Mainly on TikTok, I was like, `This looks so fire, I just need to wear it}''' (P68). Still others described copying the looks of influencers, saying things like, ``\textit{I didn't know what to do for a shirt, so I just found something from one of my YouTubers}'' (P91), and, ``\textit{Most of the skins here are from a YouTuber, or just that I found on the Internet somewhere}'' (P126). 

\subsubsection{Game Mechanics to Encourage Monetized Avatar Customization}

Game designs often include incentives that encourage players to spend more, either by offering bonus items or by creating collections that promise special rewards when completed. These strategies not only drive in-game purchases but also influence how children perceive and value their avatars.
\begin{itemize}
\item\textbf{Bonus Items and Free Incentives.} One common strategy is offering bonus items or free accessories as an incentive for making purchases. For example, one child explained how buying Robux with a gift card led to receiving a free outfit piece: ``\textit{So, the sandals came for free. If you're trying to complete a whole outfit and you buy everything, they usually give you the stuff that...} \textit{`Cause then Roblox guesses on what you're trying to wear in real life and they would just give it to you for free}'' (P68). These bonus items are often tied to popular characters or themes, making them even more appealing to children. As the participant further noted, ``\textit{If it's like a real-life character, like in a show, they'll just give it to you. 'Cause the only thing that was on the list is the hair, the Aura, the shirt, well, the shirt and the legs. There was no sandals with it, but I guess, yeah, so for the sandals, Roblox just gave it to me}'' (P68). By tying free items to beloved characters, game developers create a powerful incentive for children to make purchases of associated items, promoting their investment in their avatars.

\item\textbf{Collection Rewards.} Another monetization tactic involves encouraging players to complete collections to receive special items. This strategy leverages the desire for exclusivity and completeness, driving further spending. One participant described this mechanism in a game: ``\textit{Well, mostly because there's a thing, I forgot how to go to it, but you can get a rainbow one [car] if you collect all the colors. If I collect all of them, I can get the chromatic one, and it actually looks kinda cool}'' (P78). The promise of a unique reward for completing a set motivates children to continue purchasing items, often spending more than they initially intended. The child further explained, ``\textit{It's basically just like a spot where you can buy hats, toppers, car bodies, decals, banners. And like mystery items, you can get a whole bunch of stuff from it... You can either buy it to reveal them, or every 15 hours you could get a free drop}'' (P78).
\end{itemize}
The combination of paid and free elements, along with the potential to complete a collection to win unique items, keeps children engaged and frequently spending on avatar.

\subsubsection{The ``Wardrobe Effect'' in Avatar Customization}
We found that, despite having numerous avatars and accessories in their inventories, children tend to use one favorite avatar consistently and rarely change it, a phenomenon we call the ``\textit{wardrobe effect}.'' This behavior mirrors real-world consumer habits, where people own many outfits but only regularly wear a select few. This accumulation behavior is likely influenced by the game's monetization mechanisms, which encourage purchases through the constant introduction of new items, even if those items do not ultimately alter the player's primary avatar.

A number of children we interviewed reported that they do not frequently change their avatars, even though they have multiple options available. For many, once they find an avatar they like, they stick with it for an extended period. For instance, children mentioned, ``\textit{I just wear this all the time. I barely change}'' (P71), ``\textit{I've had this [avatar] for about a year, and I don't plan on changing it anytime soon}'' (P91), \rr{and ``\textit{I just haven't bothered to change it. I just thought they were pretty'} (P10). P79, who just showed an RA their abundant avatar collection, when asked if they have changed their avatars often, they replied, ``\textit{Actually, I've just been the Junkbot for as long as I've been playing Roblox, I've been the Junkbot and I've never changed. Like I've never been... I've never changed my avatar to like John or Lin or Oakley or anything.}'' (P79)}
These sentiments were echoed by others who similarly expressed a preference for maintaining a consistent avatar over time. 

Interestingly, even though these children do not frequently change their avatars, many still accumulate a large number of skins and accessories, \rr{and they were excited to show and explain their avatar collection and customization to the RAs during the interview}. As one participant described:

\begin{table}[H]
\centering
\begin{tabular}{r p{.72\columnwidth}}
\textbf{P130}: & 
  ``\textit{Yeah, there's a lot of stuff that I bought before that I don't like anymore}.''\\
\textbf{Interviewer}: & 
  ``\textit{So how often do you change your outfit}?''\\
\textbf{P130}: & 
  ``\textit{Not often. I've had this for a while and I don't plan on changing it}.''\\
\textbf{Interviewer}: & 
  ``\textit{So when you were on the marketplace and you bought the shirts, what about the items make you want to buy them}?''\\
\textbf{P130}: & 
  ``\textit{I mean, just because I think they look nice or I used to think they look nice}.''\\
\end{tabular}
\end{table}

\noindent This behavior reflects a consumer culture where owning more is often seen as desirable, even if the items are not regularly used.

\section{Discussion}
\subsection{Avatar Customization as a Tool for Identity Exploration}

Avatar customization in social games offers a powerful tool for children to explore and express their identities in ways that are often not possible in the real world. Unlike real-world settings, where self-expression can be limited by social norms, economic constraints, and physical characteristics, social games provide a comparatively safe and flexible space where children can experiment with different aspects of their identity, such as diverse physical traits, aesthetic preferences, and social roles. Our findings show in these virtual spaces, children can freely test different hairstyles, body types, clothing, accessories, and even non-human forms, without facing the same social or personal repercussions they might encounter in their everyday lives. \rr{This aligns with findings from research on digital identity construction, which highlight the value of anonymity and pseudonymity in fostering freer self-expression and identity experimentation in online spaces \cite{subrahmanyam2011constructing}.} The ability to quickly create and modify avatars and receive feedback from peers allows children to explore and better understand their identities through the social dynamics of gameplay. \rr{Players have abundant opportunity to experiment with the avatar-user relationship, which is fluid and changes based on the user’s goals, the particular virtual world, and the evolving nature of online identities \cite{de2012my}.}

The opportunity to explore identity is especially valuable for younger users, who are still developing their self-concept and may feel heightened social self-consciousness \cite{sebastian2008development, elkind1967egocentrism}. In social online games, children can engage in repeated iterations of avatar creation, experimenting with different versions of themselves in a relatively low-risk setting. 
\rr{The iterative exploration in digital environments enables users to navigate and reconcile multiple facets of their identity, often alternating between the actual self, the ideal self, and exploratory personas \cite{subrahmanyam2011constructing, wonica2014exploring}.} Our findings show this process allows children to explore two types of self (actual-self, ideal-self) as identified by the self-discrepancy theory \cite{higgins1987self}. A notable trend is the tendency for children to create avatars that are slightly more idealized than their real selves, incorporating traits they aspire to, such as maturity, style, or kindness. \rr{These findings echo previous research, which shows avatars offer a unique venue for exploring self and identity during adolescence \cite{wood2020me} and that ``possible selves'' \cite{markus1986possible} can be a powerful motivator, influencing behavior and choices as individuals work towards desired future identities. }

Research suggests that when an avatar closely aligns with a player's ideal self, it can enhance immersion and serve as a predictor of intrinsic motivation and positive emotions following gameplay \cite{przybylski2012ideal}. The idealized avatar may influence children's gameplay behavior since when children create and interact with an avatar that embodies their ideal self, they may begin to internalize the traits and characteristics of that avatar as part of their self-identity, according to the self-perception theory \cite{fazio2014self} and Proteus effect \cite{yee2007proteus}. For instance, if a child’s avatar is designed to be confident, kind, or courageous, they may see themselves as possessing those traits in both gameplay and real life.

While this can be empowering, there are potential risks associated with over-identification with these idealized avatars. Excessive attachment to a virtual self that differs significantly from one’s real self may contribute to negative outcomes, such as \rr{distorted self-presentation, feeling of inadequacy, comparison,} gaming disorder, addiction, social withdrawal, or reduced well-being \cite{subrahmanyam2011constructing, green2021avatar}. It may also foster a gap between the actual and the ideal self, potentially leading to feelings of inadequacy or dissatisfaction in the real world. \rr{This risk of over-identification is echoed by Wood and Szymanski who suggest that while experimentation with avatars can be positive, educators and parents need to be mindful of potential negative consequences such as excessive gaming or a blurring of lines between online and offline identities \cite{wood2020me}.}

To balance the benefits of identity exploration with the risks of over-identification, game designers should consider strategies that promote healthy engagement with avatar making and customization. For example, game mechanics could encourage players to explore rather than attaching too strongly to any single idealized version. This could include providing diverse customization options that are equally valued within the game or designing features that support diversity in avatars over time. As one child described the culture in Roblox, "Everybody that plays Roblox knows that everybody is wearing something weird," suggesting that a culture of varied and creative avatar creation can support healthy self-exploration. 

Additionally, parents, educators, and caregivers can play an important role by fostering open discussions with children about their digital identities and self-concept development. As children's experimentation with avatars may mirror their real-world identity exploration, parents can support this process by encouraging reflection and helping them articulate their experiences. They can guide children in distinguishing between experimentation in a virtual environment and their real-life self-perception,  while remaining attentive to signs of excessive attachment to avatars and exploring the underlying reasons for such behavior. \rr{By fostering digital literacy and self-awareness of their game avatars and their real-life selves, parents can mitigate risks and empower children to navigate their online and offline identities more effectively \cite{subrahmanyam2011constructing}. Previous research also argues educators can leverage avatar-related activities into the classroom to facilitate discussions about identity, perspective-taking, and communication \cite{wood2020me}.}

\subsection{Navigating Social Conformity and Self-Expression}

Social games also serve as a microcosm for the tension between social conformity and self-expression. As children transition into adolescence, peer pressure and the desire for social acceptance become increasingly significant. These dynamics are reflected in avatar customization choices, where children often balance their desire to express their individuality with the need to fit in with their peers. Our findings suggest that community building and seeking social attention are potent motivators in avatar creation, with many children choosing avatars that reflect the styles or preferences of their friends, family members, or broader social groups.

This tension between self-expression and social conformity presents both challenges and opportunities for game design. To support digital socialization while allowing for personal self-expression, game developers should create environments that celebrate both shared identity and individuality. For example, designing social avatar-making features that reward community engagement and collaboration while also providing ample customization options that allow for personal creativity and expression can help balance these competing needs. By fostering an inclusive community culture that values diverse expressions of identity, games can reduce the pressure children might feel to conform to specific social norms. \rr{Kim et al. suggest that highly customizable avatars can strengthen a user's identification with a virtual community by allowing them to express both individuality and affiliation. Conversely, games with limited customization options might inadvertently pressure players towards conformity \cite{kim2012became}.}

Game environments should also discourage negative social behaviors, such as judgment or criticism based on avatar appearance, by promoting positive interactions and providing tools for reporting or mitigating harassment. For instance, social online games could provide a wider variety of default avatars and abundant customization options from the outset to minimize the stigma associated with "newbie" avatars. \rr{Previous research finds users are highly motivated to customize their avatars to avoid the stigma of being newcomers \cite{ducheneaut2009body}.} Creating a safe and inclusive digital social space where children can freely explore their identities is crucial for supporting the development of social-emotional skills and self-expression in the virtual world.

\subsection{Cultural Consumerism and Social Media Influence}

Research has also shown that digital environments, like social media, can amplify social self-consciousness among children and adolescents. Adolescents' need for social validation and their heightened self-awareness is often intensified by online platforms where they are exposed to constant feedback and comparison \cite{boursier2020objectified}. 
Our results show that cultural consumerism has a significant effect on children's avatar creation and purchasing behavior. Our results especially show that digital media consumption, such as movies and TV shows, and social media, such as TikTok, influence children's avatar choices. As cultural consumerism influences our buying behavior in the world, we discovered children's avatar making also reflects this phenomenon that they own a collection of avatars and accessories but only consistently wear one, \rr{which we termed the ``wardrobe effect.'' Just as a physical wardrobe can represent a collection of potential selves or aspirational identities, the accumulation of avatars might serve a similar function in games. Children might acquire different avatars to align with specific moods, social groups, or desired personas, even if they ultimately gravitate towards a limited selection for regular use. For some, avatar ownership and customization provide value and improve gaming experience, even if the avatars aren't used in-game.  In these cases, avatar creation is more about personal satisfaction than external presentation. In addition, design factors could contribute to this wardrobe effect, for instance, game design mechanics, such as loot boxes or gacha systems, can incentivize the collection of avatars and accessories, even if those items are rarely used in gameplay. Additionally, children gravitate to colorful, whimsical, and fantastical avatars and customizations \cite{kriglstein2013study}, which can be leveraged by cultural consumerism and game companies. This raises ethical concerns about the potential for these mechanics to exploit children's desire for novelty and social validation.}

This game design pattern raises concerns about overconsumption and potentially nudges young players to overspend. To mitigate these negative effects, game designers can employ strategies that promote creativity over-commercialization, such as providing robust, accessible tools for avatar customization. Another approach game designers can employ is to encourage meaningful engagement over mere accumulation by rewarding the creative use of existing avatar options. Lastly, the recent introduction of generative AI technologies holds significant potential in this regard. For instance, Roblox announced a new tool that will enable easy creation of custom avatars from images, which could be promising to democratize avatar creation and support children's creativity and self-expression \cite{robloxGenAI}. Companies can align their monetization strategies with supporting children's creativity and self-expression by offering advanced customization tools at low or no cost, adopting transparent pricing models, and prioritizing ethical considerations in supporting children.

\section{Limitation and Future Work}
Our study involved 48 participants aged 8 to 13, which may not represent the full range of children's age groups. In addition, despite our efforts to recruit a diverse sample in terms of ethnicity, our participant population lacked diversity, with the white children being our most recruited participants. Additionally, we primarily focused on three of the most popular social games—Roblox, Minecraft, and Fortnite—which may limit the generalizability of our findings to other social gaming platforms. Furthermore, we conducted interviews and observations of children's gaming behavior in only one session. Observing children's avatar modifications over time would be a valuable direction for future research, particularly as they grow and technology evolves. \rr{Our data doesn't explain why some children modify avatars more frequently or if game design plays a key role. Future research should explore the primary and secondary causes of this effect.} 

\textcolor{orange}{Additionally, while our study surfaced the ``wardrobe effect,'' where children collect numerous avatars or accessories but primarily use a single favorite, this topic was not a designed study focus. Future research could examine whether this behavior is linked to offline consumer habits and, if so, explore the underlying causes. Such studies could also investigate whether this tendency is influenced by the monetization strategies employed by games. Finally, future research should consider isolating the effects of monetization and cultural consumerism on children's avatar decision-making.}

\section{Conclusion}
Our study shows children aged 8 to 13 engaging with avatar creation and customization in social online games like Roblox, Minecraft, and Fortnite. Through semi-structured interviews and gameplay observations with 48 participants, we found that children's motivations for avatar customization are multifaceted, encompassing self-representation, experimentation with alter egos, social connection, and perceived in-game advantages. In addition, we found designed monetization mechanisms influence children's avatar making. The ``wardrobe effect'' highlights a pattern of overconsumption influenced by cultural consumerism and social media, where children accumulate numerous virtual items but consistently use only a favored few. 

These findings contribute to a deeper understanding of how online games serve as platforms for identity exploration among young players. By recognizing the impact of internal motivation and external influences, such as game design and monetization mechanisms on children, we underscore the need for ethical and thoughtful social online game design.

\begin{acks}
We would like to thank Kaitlin Tiches for helping us find relevant literature. The Digital Wellness Lab at Boston Children’s Hospital is supported by Amazon Kids, Discord, Meta, Pinwheel, Pinterest, Point 32 Health: Harvard Pilgrim Health Care/Tufts Health Plan, Roblox, Snap Inc., TikTok, Trend Micro Cares, and Twitch. These organizations provided unrestricted research funding and were not involved in the design, analysis, or reporting of this research.

Alexis Hiniker and Jenny Radesky are special government employees for the Federal Trade Commission. The content expressed in this manuscript does not reflect the views of the Commission or any of the Commissioners. 

\end{acks}

\bibliographystyle{ACM-Reference-Format}
\bibliography{Avatar}


\appendix
\section{Appendix}
\label{Appendix:A}

\begin{figure*}[!htb] 
    \centering
    \includegraphics[width=0.5\linewidth]{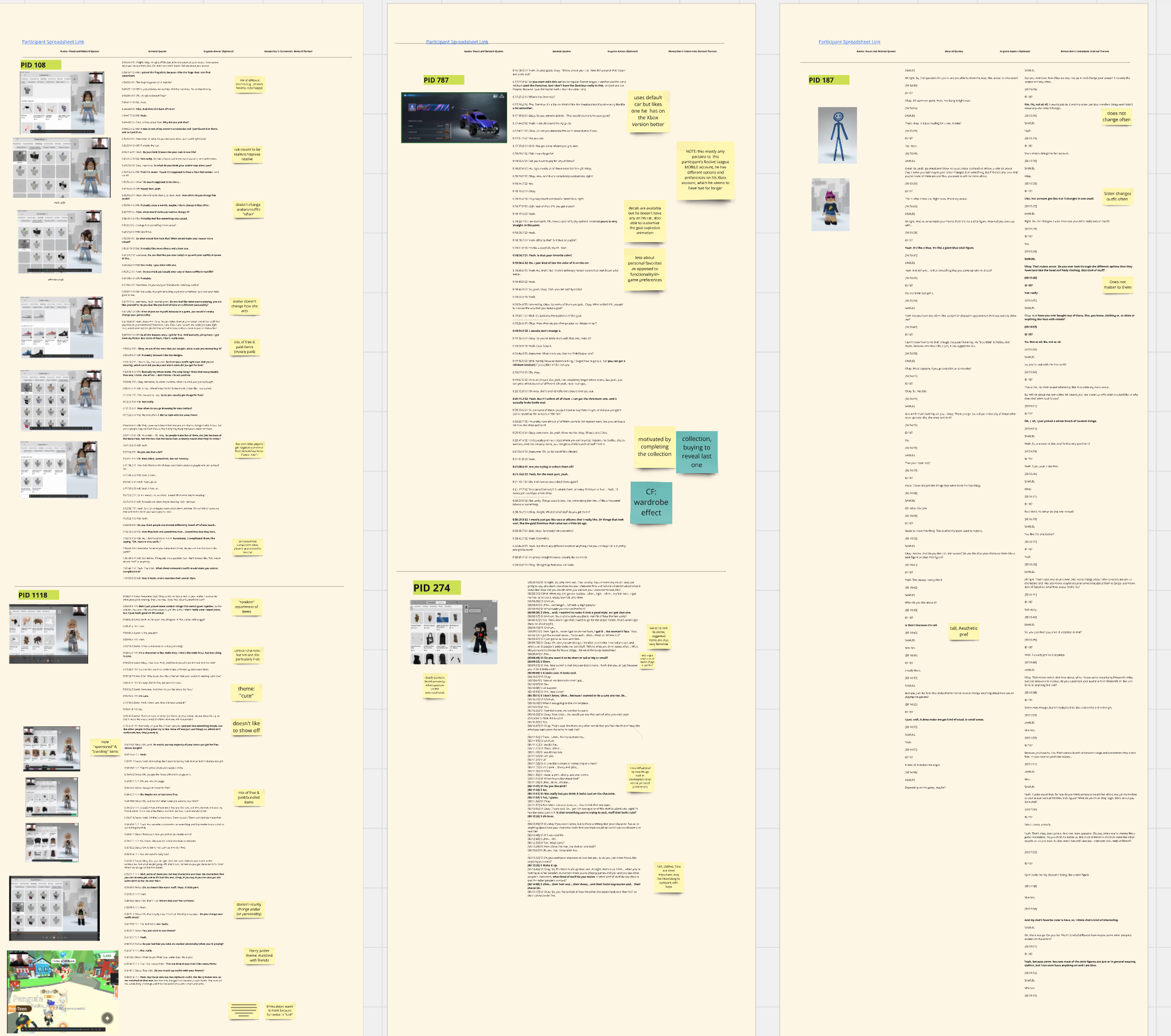} 
  \caption{ \rr{Extracted interview transcript data and visuals from recordings, accompanied by research assistants' notes and reflections on participants' avatars during the analysis process.}}
    \label{fig:miro coding}
\end{figure*}

\begin{figure*}[!htb] 
    \centering
    \includegraphics[width=0.5\textwidth]{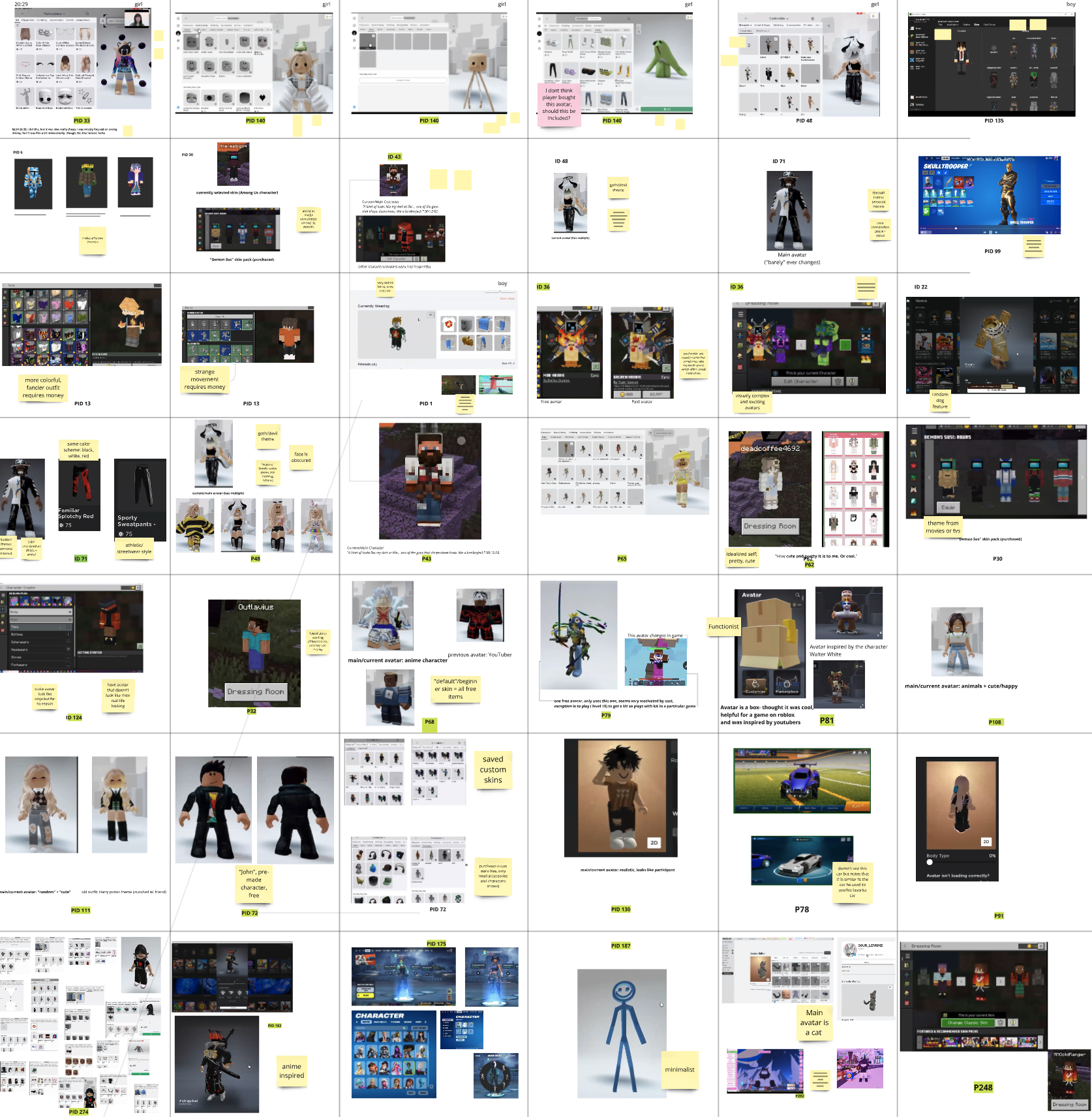} 
    \caption{\rr{An image board showing the visual data analysis process, used to explore participants' avatar characteristics and the underlying motivation behind children's avatar choice.}}
    \label{fig:coding image board}
\end{figure*}

\begin{figure*}[!htb] 
    \centering
    \includegraphics[width=0.8\textwidth]{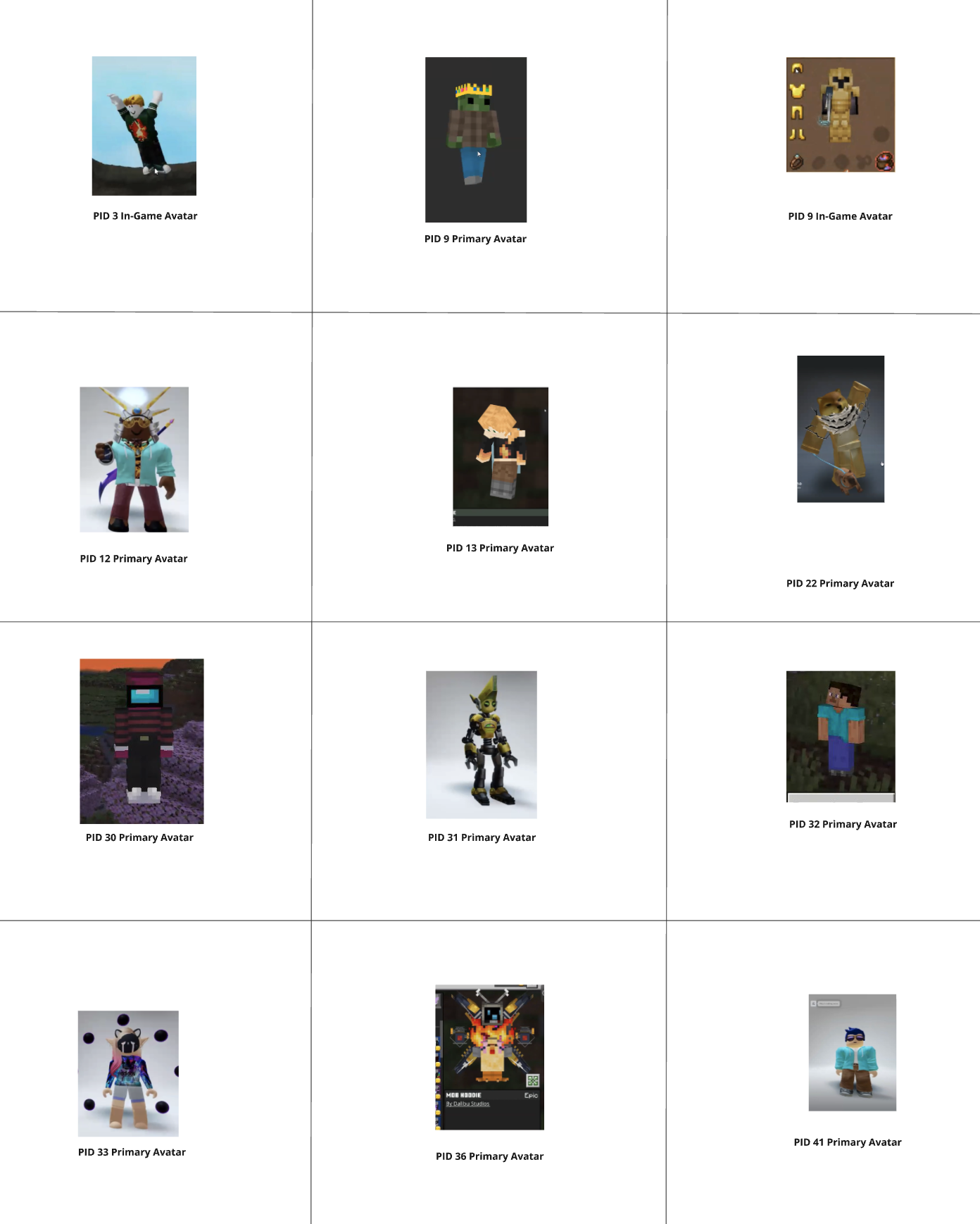} 
  \caption{\rr{A sample of participants' avatar visuals. The full dataset is made available on the ACM Library.}}
    \label{fig:avatar sample}
\end{figure*}

\end{document}